%Paper: hep-th/9303050
%From: shin@SHIN.kyunghee.ac.kr
%Date: Tue, 9 Mar 93 18:53:13 KST

\input phyzzx.tex
\vskip-10pt
\hfill
{KHTP-93-01\par}
\vskip-15pt
\hfill
{SNUCTP-93-05\par}
\title{\seventeenbf Conformal Turbulence with Boundary}
\author{B.K. Chung, Soonkeon Nam,
\foot{E-mail address: nam@nms.kyunghee.ac.kr}
Q-Han Park,
\foot{E-mail address: qpark@nms.kyunghee.ac.kr}
and H.J. Shin
\foot{E-mail address: shin@SHIN.kyunghee.ac.kr} }
\address{\it  Department of Physics\break
          Kyung Hee University\break
          Seoul, 130-701, Korea}
\abstract{
Based upon the formalism of conformal field theory with a boundary,
we give a description of
the boundary effect on fully developed two dimensional
turbulence. Exact one and two point velocity correlation
functions and energy
power spectrum confined in the upper half plane
are obtained using the image  method.
This result enables us to address the infrared problem of the
theory of conformal turbulence.}
\endpage
%%%%%%%%%%%%%%%%%%%%%%%%%%%%%%%%%%%%%%%%%%%%%%%%%%% %%%%%%%%%%%%%%%%%%%%%%%%%
\def\zb{{\overline z}}

%%%%%%%%%%%%%%%%%%%%%%%%%%%%%%%%%%%%%%%%%%%%%%%%%%% %%%%%%%%%%%%%%%%%%%%%%%%

\REF\turb{See for example,
L.D. Landau and E.M. Lifshitz, Fluid Mechanics, Pergamon, Oxford (1984).}
\REF\book{A.S. Monin and A.M. Yaglom, Statistical Fluid
Mechanics, The MIT Press, Cambridge (1975).}
\REF\Migdal{A.A. Migdal, Mod. Phys. Lett. {\bf A6} (1991) 1023.}
\REF\PolyI{A.M. Polyakov, Princeton Univ. preprint PUPT-1341,
hep-th/9209046.}
\REF\PolyII{A.M. Polyakov, Princeton Univ. preprint PUPT-1369,
hep-th/9212145.}
\REF\turbexp{B. Legras, P. Santangelo, and R. Benzi, Europhys. Lett. {\bf 5}
(1988)
37. }
\REF\Kol{A.N. Kolmogorov, J. Fluid Mech. {\bf 13} (1962) 82.}
\REF\CardyII{J.L. Cardy, Phys. Rev. Lett. {\bf 54} (1985) 1354.}
\REF\Kraichnan{R. Kraichnan, Phys. Fluid, {\bf 10} (1967) 1417.}
\REF\Ferretti{G. Ferretti and Z. Yang, Rochester preprint UR-1296,
hep-th/9212021.}
\REF\Low{D.A. Lowe, Princeton University preprint PUPT-1362, hep-th/9212019.}
\REF\Falkovich{G. Falkovich and A. Hanany, WIS-92/88/Nov-PH, hep-th/9212015. }
\REF\Matsuo{Y. Matsuo, Univ. of Tokyo preprint UT-620, hep-th/9212010.}
\REF\CardyI{J.L. Cardy, Nucl. Phys. {\bf B240} [FS12] (1984) 514.}

Turbulence arises in many physical systems, which obey
simple but nonlinear equations of motions.\refmark\turb\
The statistical nature of turbulence together with its nonlinear,
nonequilibrium structures seem to require certain
non-unitary quantum field theories for the theory of turbulence.\refmark\book\
The appearance of divergences in convensional perturbative
approach\refmark\Migdal\ suggest the use of renormalization group technique.
However, so far this has been unsuccessful leaving turbulence still
an outstanding problem in theoretical physics.

Recently, Polyakov\refmark{\PolyI,\PolyII}\ has proposed a new approach
based on the assumption of conformal invariance in the inertial range of fully
developed two dimensional turbulence. He argued that certain non-unitary
conformal field
theories (CFT's) provide exact solutions of the fundamental equations of
turbulence,
which are  consistent with the known energy spectrum.\refmark\turbexp\
However, Polyakov's solutions in terms of correlators of non-unitary CFT are
plagued with infrared(IR) divergences and also highly dependent on the
arbitrary vacuum expectation values of operators in the theory.  Instead of
taking the IR
problem as a weakness of his approach, Polyakov\refmark{\PolyI,\PolyII}\
has conjectured that the resolution of the problem is actually the most
important part of
the theory.

In this letter, we introduce a physical boundary to a turbulent system in order
to regulate
the IR divergence. Using an `image method', we determine exact one and two
point velocity
correlation functions and energy power spectrum which depend on the
distance from the boundary. Away from the boundary, these functions quickly
become isotropic which is in accordance with the Kolmogorov's local isotropy
assumption.\refmark{\Kol,\book}\  This allows us to use the boundary not
only for handling the IR divergences, but also for an effective tool to
understand the IR behavior of the theory.  We attempt to give a
physical interpretation to these results.

In two dimensions, the fluid motion is governed by the Navier-Stokes
equation:  $$\dot\omega+\epsilon_{\alpha\beta}\partial_{\alpha}\psi
\partial_{\beta}\partial^{2}\psi = \nu\partial^{2} \omega
+{\rm stirring\  force},\quad\quad (\alpha,\beta = 1,2)\eqn\NS$$
where $\omega$ and $\psi$ denote the vorticity and the stream
function respectively and $\nu$ is the viscosity. Stream function
is related to the vorticity and  velocity through $\omega =\partial^2 \psi$
and  $v_\alpha = \epsilon_{\alpha\beta}\partial_{\beta}\psi$.
A typical stirring force is provided by letting fluid flow through a grid
consisting of very thin bars.  For fluid with a sufficiently large Reynolds
number, fully developed turbulence emerges 30 to 40 times the grid bar spacing
away, where the stirring force is absent.\refmark\book\ The boundary which we
use in this paper is parallel to the direction of the incoming flow.

In the following, we first review the case without boundary. Polyakov's CFT
approach is to identify the stream function $\psi$ with a primary field of a
certain minimal CFT.  In the static and inviscid case, the Navier-Stokes
equation reduces to  $ ``\epsilon_{\alpha\beta}\partial_{\alpha}\psi
\partial_{\beta}\partial^{2}\psi "=0$. This equation has meaning only when we
exercise care in defining  the operator product at the coinciding points,
e.g., by using  the point splitting method,
 $$``\epsilon_{\alpha\beta}\partial_{\alpha}\psi (z)
\partial_{\beta}\partial^{2}
\psi(z)" = \lim _{a\rightarrow 0} \epsilon_{\alpha\beta}\partial_{\alpha}\psi
(z+a)
\partial_{\beta}\partial^{2}\psi (z),\eqn\ope$$
where $\lim$ implies angle averaging. In CFT, the operator product expansion
(OPE) of $\psi$ has the following  structure:
$$\psi (z+a)\psi (z) = (a\overline{a})^{\Delta_{\phi}-2\Delta_{\psi}}
\{ \phi (z) +{\rm descendents} \}+\cdots ,\eqn\opeII$$
where $\phi$ is the minimal dimension operator of  dimension $\Delta_\phi$
which is most  relevant as $a\rightarrow 0$.  This leads to
$$``\epsilon_{\alpha\beta}\partial_{\alpha}\psi (z)
\partial_{\beta}\partial^{2}\psi (z)" \sim {\lim_{a\rightarrow 0}}
(a\overline{a})^{\Delta_{\phi}-2\Delta_{\psi}} [L_{-2}{\overline
L}^{2}_{-1}-{\overline L}_{-2}L_{-1}^{2}]\phi(z),\eqn\confns$$
with $L_{-n}$ being Virasoro generators. The right hand side of Eq.\confns \
vanishes, i.e., the Navier-Stokes equation is satisfied, for the following
two cases: First is  when $[L_{-2}{\overline L}^{2}_{-1}-{\overline
L}_{-2}L_{-1}^{2}]\phi $$= 0$, which for example is satified when $\phi$ is
degenerate on the level two. The simplest concrete example of this is the
minimal
model of $p=2, q=5$, which we denote $\cal{M}_{\rm (2,5)}$, the critical
Yang-Lee edge singularity.\refmark\CardyII\  Second case is when
$\Delta _{\phi}>2\Delta_{\psi}$. This together with a further restriction
coming from the constant enstrophy flux condition\refmark\Kraichnan\
 has led Polyakov and others find series of exact solutions of conformal
turbulence.\refmark{\PolyI,\PolyII,\Ferretti -\Matsuo}\
The first case of ${\cal{M}}_{\rm (2,5)}$ does not have constant
enstrophy flux, thus  requires a subtle balance between vorticity and
energy injection.\refmark\PolyI\

In order to calculate the energy spectrum $E(k)$,
we need the two point function of $\psi$;
$$\left\langle\psi(z_{1})\psi (z_{2})\right\rangle\sim  {\langle I \rangle
\over
|z_{1}-z_{2}|^{4\Delta_{\psi}}} + {\langle \phi (z_{2})\rangle \over
|z_{1}-z_{2}|^{4\Delta_{\psi}-2\Delta_{\phi}}}+ \cdots,\eqn\tpf $$
which results in the power law behavior
$$ E(k)\sim k^{\alpha},\eqn\energy$$
where $\alpha=4\Delta_{\psi}+1$ for $\langle \phi\rangle=0$,
and $\alpha=4\Delta_{\psi}-2\Delta_{\phi}+1$ for $\langle \phi\rangle\neq 0$.
Unlike in unitary cases,  one point function  $\langle \phi\rangle$ is
not necessarily zero in non-unitary theories,
and as we will see in the following, $\langle \phi\rangle$ can be determined
by boundary condition.

In this letter, we will focus on $\cal{M}_{\rm (2,5)}$ minimal model
restricted to the upper half plane $(y>0)$ to investigate boundary effects on
fully developed turbulence explicitly.
One way to see the effect of a boundary on a turbulent flow
would be to imagine that each eddy in turbulent flow has
an image eddy on the other side of the boundary, similar to image
charges of electrostatics near a conductor. This is
consistent with the fact that vertical component of fluid velocity
vector vanishes at the boundary. Furthermore, this picture enables
us to extend Polyakov's CFT approach to turbulence to the case with boundary.

Cardy\refmark\CardyI\  introduced CFT with boundary utilizing the image method
to describe  surface critical behavior in two dimensions.
He has shown that the restricted conformal transformation
which preserves boundary conditions can be achieved by
extending the energy momentum tensor $T(z)$ analytically in the lower
half plane through $T(z)={\overline T}(\zb)$, when $y=Im\ z<0$.
One consequence of it is that for the
degenerate conformal field theories, the $n$-point function
$\langle \phi (z_{1},\zb_{1})\cdots\phi (z_{n},\zb_{n})
\rangle_{b}$ can be obtained systematically from
the bulk $2n$-point function, where $\langle\cdots\rangle_{b}$ denotes
correlation functions in the upper half plane.
This is so because the $n$-point function $\langle \phi (z_{1},\zb_{1}) \cdots
\phi (z_{n},\zb_{n})\rangle_{b}$
satisfies the same differential equation as does the
bulk $2n$ point function consisting of charges in the upper
half plane as well as their images in the lower half plane.

${\cal M}_{\rm (2,5)}$ minimal  model has only one nontrivial primary field
$\phi_{(1,2)}$,
of conformal dimension $\Delta_{(1,2)}=-1/5$, whose normalization is such
that $\langle\phi_{(1,2)}(z_{1})\phi_{(1,2)}(z_{2})\rangle
=|z_{1}-z_{2}|^{4/5}$.
Here we identify $i$ times the primary field
$\phi_{(1,2)}$ as the stream function $\psi$.  The factor $i$ is required  for
the following
reason; if we neglect the factor $i$, we can not simultaneously have a real
one point function and a positive two point function of velocity, which are
necessary in order to be physical.

Following the Cardy's image method, we obtain the one point function of $\psi$
with boundary from a two point function of $\psi(z)$ and its image
$\psi(\zb)$ across the boundary such that
$$\langle\psi (z,\zb)\rangle_{b} = \langle\psi (x,y)\rangle_{b} =
d_\psi  y^{2/5},\  (z=x+iy, y>0),\eqn\tpfwb$$
where $d_{\psi}$ is an arbitrary real constant.
In the unitary case with boundary, $d_{\psi}$
is fixed by the asymptotic behavior of two point fuction:
$\langle\psi(x_{1},y_{1})\psi(x_{2},y_{2})\rangle_{b}
\rightarrow\langle\psi(y_{1})\rangle_{b}\langle\psi(y_{2})\rangle_{b}$ as
$|x_{1}-x_{2}|\rightarrow \infty$,
whereas in the non-unitary case $d_{\psi}$ is in general
arbitrary. The mean velocity can be obtained from the one point function of
$\psi$
such that $$\eqalign{& \langle v_{x}(x,y)\rangle_{b}= \langle\partial_{y}
\psi(x,y)\rangle_{b}= {2\over 5} d_{\psi} y^{-3/5},\cr
&\langle v_{y}(x,y)\rangle_{b}= -\langle\partial_{x}
\psi(x,y)\rangle_{b}=0.\cr}\eqn\opfwb$$ We interpret this result as describing
a system
with a net drift velocity profile of turbulent flow  parallel to the boundary.
The drift
velocity in Eq.\opfwb \ goes to infinity as $y\rightarrow 0$, so it should be
rounded off
over a certain  distance from the boundary.

In order to calculate the energy spectrum of
the system, we need the two point function of
velocity. Again, this can be obtained from the
bulk four point function of $\psi$
involving two in the upper half plane and two images in the lower half.
For $\cal{M}_{\rm(2,5)}$, which is degenerate on
the level two, the bulk four point function
satisfies a second order differential equation.
Solutions of the differential equation can be put in terms of linear
combinations of hypergeometric functions of variable
$\xi=-{|z_{1}-z_{2}|^{2}\over 4y_{1}y_{2}}$:
$$\langle \psi (z_1,\zb_1) \psi(z_2,\zb_2)\rangle_{b}
= \left({z_{13}z_{24}\over z_{12}z_{23}z_{34}z_{14}}\right)^{-2/5}
\left(aF_{1}(\xi)+bF_{2}(\xi)\right),\eqn\tpfwbii$$
where we use the notation  $z_{3}=\zb_{1}, \ z_{4}=\zb_{2}$,
$z_{ij}=z_{i}-z_{j}$,
and $F_{1}= \ _{2}F_{1}({3\over 5},{4\over 5};{6\over 5};\xi)$,
$F_{2}=\xi^{-1/5}\ _{2}F_{1}({3\over 5},{2\over 5};{4\over 5};\xi)$.
$a$ and $b$ are constants which can be determined
by looking at the limiting behavior of the correlator as
$|z_{1}-z_{2}|\rightarrow 0$.
In this limit, the right hand side of Eq.\tpfwbii \ becomes
$$\langle \psi (z_1,\zb_1) \psi(z_2,\zb_2)\rangle_{b}
\rightarrow |z_{1}-z_{2}|^{4/5}\left[a+b\left(-{|z_{1}-z_{2}|^{2}\over
4y^{2}}\right)^{-1/5} \right],\eqn\limtwo$$
where $y=(y_{1}+y_{2})/2$. On the other hand,
the OPE of $\psi(z_{1}) \psi(z_{2})$ as $z_{1}\rightarrow z_{2}$ gives
$${C_{\psi\psi I}\over |z_1-z_2|^{4\Delta_{\psi}}}\langle I\rangle_{b}+
{C_{\psi\psi\psi}\over |z_1-z_2|^{4\Delta_{\psi}-2\Delta_{\psi}}}
\langle\psi\rangle_{b}=  -| z_1-z_2|^{4/5}+|z_1-z_2|^{2/5}
C_{\psi\psi\psi}d_{\psi}y^{2/5}.\eqn\popo$$
$C_{\psi\psi I}$ and $C_{\psi\psi\psi}$ are structure functions which
are determined from the three point function of the bulk theory.
Explicit calculation shows that
$C_{\psi\psi I}=-1$ and $C_{\psi\psi\psi}=|C|^{1/2}$,
where $C=-{\Gamma(6/5)^{2}\Gamma(1/5)\Gamma(2/5)\over
\Gamma(3/5)\Gamma(4/5)^{3}}$.\refmark\CardyII\
Comparing Eq.\limtwo \ and Eq.\popo, we get $a=-1$ and $b=2^{-2/5}d_{\psi}
|C|^{1/2}$.

The correlators of velocity components can be
obtained by differentiating the stream function
correlator, for example, $\langle v_{y}(z_{1})v_{y}(z_{2})\rangle_{b} =
\partial_{x_{1}}\partial_{x_{2}}\langle \psi(z_{1})\psi(z_{2})\rangle_{b}$.
In order to obtain the energy spectrum $E(k)$ for $k\gg k_{0}$ where
$k_{0}$ is the characteristic input wavenumber,
we consider the form of velocity correlators in the small distance limit.
When $x_1=x_2$ and
$y_{1}=y+\epsilon/2$, $y_{2}=y-\epsilon/2$ ($\epsilon \ll y$),
the velocity correlator is given by
$$\eqalign{ &
\langle v_{y}(x_{1},y_{1}) v_{y}(x_{2},y_{2})\rangle_{b} =
{4\over 5} \epsilon^{-6/5}\left(1-{3\over 440}{\epsilon^{4}\over y^{4}}
-{7\over 7040}{\epsilon^{6}\over y^{6}} + \cdots\right) \cr
& \quad\quad \ \
+ {2\over 5}d_{\psi}|C|^{1/2}\epsilon^{-8/5} y^{2/5} \left(1+{1\over
10} {\epsilon^{2}\over y^{2}}-{7\over 600}{\epsilon^{4}\over y^{4}}-{3\over
2000}
{\epsilon^{6}\over y^{6}}+ \cdots\right).\cr}\eqn\velcorr$$
When the mean velocity is zero ($d_{\psi}=0$) and when moved far away from the
boundary,
i.e., when $y\gg 1$, the isotropic bulk two point velocity correlation function
is
recovered. Similarly, we have
$$\eqalign{ &
\langle v_{x}(x_{1},y_{1}) v_{x}(x_{2},y_{2})\rangle_{b} =
-{4\over 25} \epsilon^{-6/5}\left(1+{57\over 440}{\epsilon^{4}\over y^{4}}
+{243\over 7040}{\epsilon^{6}\over y^{6}} + \cdots\right) \cr
& \quad\quad \ \
- {6\over 25}d_{\psi}|C|^{1/2}\epsilon^{-8/5} y^{2/5} \left(1+{1\over 10}
{\epsilon^{2}\over y^{2}}+{179\over 1800}{\epsilon^{4}\over y^{4}}+{47\over
2000}
{\epsilon^{6}\over y^{6}}+ \cdots\right),\cr}\eqn\velcorrii$$
and also $\langle v_{x}(x_{1},y_{1}) v_{y}(x_{2},y_{2})\rangle_{b} =0$.
It is straighforward to show that the leading order terms in Eq.\velcorr \ and
Eq.\velcorrii \ are isotropic in the sense that they do not depend on the
limiting procedure of $|z_{1}- z_{2}|\rightarrow 0$.

The energy density at point $(x,y)$ is given by
${1\over 2} \langle v_{\alpha}^{2}(x,y)\rangle_{b} = \int dk
E_{(x,y)}(k)$.  In the momentum space, the energy spectrum $E_{(x,y)}(k)$
is given in terms of velocity correlators:
$$E_{(x,y)}(k)\sim {1\over 8\pi^{2}}
\int d^{2}x' e^{ik_{\alpha}(x_{\alpha}-x'_{\alpha})}\langle
v_{\beta}(x',y') v_{\beta}(x,y) \rangle_{b}.\eqn\ednk$$
In  the homogeneous case,  $E_{(x,y)}(k)$ does not depend on the position
$(x,y)$
which, however in the presence of boundary, is a function of the position.
{}From Eqs.\velcorr  -\ednk \ , we get
$$E_{(x,y)}(k) \sim \pi \left(A k^{1/5}+B d_{\psi}y^{2/5} k^{3/5}\right),
\eqn\fff$$
where $A = 2^{-1/5}{16\over 25} {\Gamma (2/5)\over \Gamma(3/5)}$
and $B = 2^{-3/5}{4\over 25} |C|^{1/2} {\Gamma (1/5)\over \Gamma(4/5)}$.
It is remarkable that the boundary effect explicitly shows up through
the one point function of the stream function
$\langle \psi (z)\rangle_{b} = d_{\psi} y^{2/5}$.
In general, when $\langle\phi\rangle$ in Eq.\tpf \ does not vanish,
Polyakov argued that the energy spectrum is
described by $E(k) \sim k^{4\Delta_{\psi}-2\Delta_{\phi}+1}$ while for
$\langle\phi
\rangle_{b}=0$, $E(k) \sim k^{4\Delta_{\psi}+1}$.
Above two cases emerge in a unified form in our approach, and from this we see
that the spectrum near the boundary is dominated by $E(k)\sim
k^{4\Delta_{\psi}+1}$ and as the distance from the boundary increases, the
$E(k)\sim k^{4\Delta_{\psi}-2\Delta_{\phi}+1}$ behavior becomes more and more
dominant. This is one of our main result, which seems to be a salient boundary
effect
in any CFT approach to turbulence. It would be very interesting to verify such
a
phenomena experimentally in real turbulent systems.

One of the basic assumptions in turbulence theory
is the Kolmogorov's idea of local isotropy, that is, small-scale disturbances
in turbulence with sufficiently large Reynolds number
can be regarded as isotropic, regardless of large scale anisotropy.
This is consistent with our CFT approach. In our case, large scale
anisotropy is introduced through the boundary. Nevertheless small scale
fluctuations become isotropic quickly as we move away from the
boundary. In order to see this explicitly, we consider the velocity two point
function  of Eq.\velcorr \  rotated by 90 degrees with respect to the boundary.
This is given by
$$\eqalign{ & \langle v_{x}(x_{1},y_{1}) v_{x}(x_{2},y_{2})\rangle_{b} =
{4\over 5} \epsilon^{-6/5}\left(1-{3\over 440}{\epsilon^{4}\over y^{4}}
+{57\over 7040}{\epsilon^{6}\over y^{6}} + \cdots\right) \cr
& \quad\quad \ \  + {2\over 5}d_{\psi}|C|^{1/2}
 \epsilon^{-8/ 5} y^{2/ 5} \left(1+{1\over 20}
{\epsilon^{2}\over y^{2}}-{23\over 300}{\epsilon^{4}\over y^{4}}+{9\over 250}
{\epsilon^{6}\over y^{6}}+ \cdots\right),\cr}\eqn\velcorrel $$
where the limit is now such that $y_{1}=y_{2}=y$ and $x_{1}=x_{2}+\epsilon$.
Comparing this with Eq.\velcorr , we see that the isotropy is broken only
at the sixth order of $\epsilon/ y$
in $d_{\psi}$ independent terms, and  at the second order
in $d_{\psi}$ dependent terms.

In this letter, we have determined one and two point velocity
correlation functions by introducing a boundary,  which incorporates
the large scale structure into the scheme of CFT approach to turbulence.
In his application of CFT to the turbulence problem,
Polyakov\refmark{\PolyI,\PolyII}\ introduced an IR cut-off by adding $\delta$
functions
in the  momentum space.
In our approach, a more realistic IR cut-off is introduced through a boundary.
Although our model defined on a half plane still has an IR
divergence problem, this can easily be remedied by employing more complicated
boundaries.  Details about other types of boundary
and also different minimal CFT's leading to more realistic cases will be
considered elsewhere.

\noindent
{\bf Acknowledgements} \par
We would like to thank Yup Kim for discussions.
This work was supported in part by the program of
Basic Science Research, Ministry of Education,
and by Korea Science and Engineering Foundation.
\endpage
\singlespace
\refout
\end